%% file: main.tex
\documentclass[conference]{ieeeconf}
\usepackage{cite}
\usepackage{amsmath,amssymb,amsfonts}
\usepackage{algorithmic}
\usepackage{graphicx}
\usepackage{subcaption}
\usepackage{textcomp}
\usepackage{multirow}
\usepackage[ruled]{algorithm2e}
\usepackage{mathrsfs} 

\usepackage[dvipsnames]{xcolor}
\def\BibTeX{{\rm B\kern-.05em{\sc i\kern-.025em b}\kern-.08em
    T\kern-.1667em\lower.7ex\hbox{E}\kern-.125emX}}

\input{utils.tex}

\usepackage[utf8]{inputenc} 

\begin{document}

\title{\LARGE Experimental Validation of User Experience-focused Dynamic \\ Onboard Service Orchestration for Software Defined Vehicles}


\author{
    Pierre Laclau$^{1,2}$, Stéphane Bonnet$^1$, Bertrand Ducourthial$^1$, Trista Lin$^2$ and Xiaoting Li$^2$ \\
    \small{$^1$Heudiasyc, CNRS, Université de Technologie de Compiègne, France \{firstname.lastname@utc.fr\}} \\
    \small{$^2$Stellantis, Poissy, France \{firstname.lastname@stellantis.com\}}
}

\maketitle

\input{sections/00_abstract.tex}

\input{sections/01_introduction.tex}
\input{sections/02_background.tex}
\input{sections/03_problem.tex}
\input{sections/04_setup.tex}
\input{sections/05_results.tex}
\input{sections/10_conclusion.tex}

\bibliographystyle{IEEEtran}
\bibliography{references}

\end{document}

%% file: utils.tex
\newcommand{\mr}[2]{\multirow{#1}{*}{#2}}

\usepackage{array}
\newcolumntype{x}[1]{>{\centering\arraybackslash\hspace{0pt}}p{#1}}

\newcommand{\probP}{\text{I\kern-0.15em P}}


%% file: sections/00_abstract.tex
\begin{abstract}
    In response to the growing need for dynamic software features in automobiles, Software Defined Vehicles (SDVs) have emerged as a promising solution. They integrate dynamic onboard service management to handle the large variety of user-requested services during vehicle operation. Allocating onboard resources efficiently in this setting is a challenging task, as it requires a balance between maximizing user experience and guaranteeing mixed-criticality Quality-of-Service (QoS) network requirements. Our previous research introduced a dynamic resource-based onboard service orchestration algorithm. This algorithm considers real-time in-vehicle and V2X network health, along with onboard resource constraints, to globally select degraded modes for onboard applications. It maximizes the overall user experience at all times while being embeddable onboard for on-the-fly decision-making. A key enabler of this approach is the introduction of the Automotive eXperience Integrity Level (AXIL), a metric expressing runtime priority for non-safety-critical applications. While initial simulation results demonstrated the algorithm's effectiveness, a comprehensive performance assessment would greatly contribute in validating its industrial feasibility. In this current work, we present experimental results obtained from a dedicated test bench. These results illustrate, validate, and assess the practicality of our proposed solution, providing a solid foundation for the continued advancement of dynamic onboard service orchestration in SDVs.
\end{abstract}

%% file: sections/01_introduction.tex
\section{Introduction}

As the automotive landscape evolves, the demand for increasingly complex and connected features such as automated driving, advanced infotainment, and Vehicle-to-Everything (V2X) cooperative services is on the rise~\cite{buckl_software_2012}. Currently, each of these features is implemented as static, monolithic systems that are difficult to update and maintain \cite{askaripoor_ee_2022}. Onboard resources are statically allocated before manufacturing and occasionally tweaked through Over-the-Air (OTA) updates, which are limited in scope and frequency \cite{jiang_vehicle_2024,ayres_continuous_2021}. With app stores and feature sets growing, current vehicles are unable to keep pace with user expectations \cite{laclau_predictive_2023}.

Traditionally, vehicle differentiation was primarily based on static hardware variants, involving the (de)activation of features in pre-provisioned software components and network allocations \cite{schindewolf_comparison_2022}. However, the introduction of thousands of optional applications in upcoming app stores requires the implementation of a dynamic mechanism capable of adjusting network and computing allocations in response to the varying preferences unique to each user and resources.

In the future, dynamic service management features are required to extend vehicle capabilities by allocating resources and executing services depending on the current context. Software Defined Vehicles (SDVs) \cite{haeberle_softwarization_2020} are emerging as a promising solution as they have the potential to offer dynamic orchestration. They aim to (1) concentrate the execution of features in fewer more powerful Electronic Control Units~(ECUs), (2) implement virtualization techniques to enable dynamic resource allocation \cite{bandur_making_2021}, and (3) set Ethernet as the main network backbone offering high flexibility and reconfigurability with service-oriented communications \cite{hackel_secure_2022-1}. 

However, the dynamic nature of resources poses several challenges, including changes in reserved safety-critical resources based on orchestration, fluctuating network resources in V2X and cloud connectivity, as well as changing user requests that affect resource allocations. Hence, the vehicle must adapt to these changes while respecting the Service Level Agreements (SLAs), which are Quality of Experience (QoE) constraints for users defined by the OEMs, and the user preferences \cite{taylor_digital_2024}. In this context, the vehicle must dynamically orchestrate services while respecting the SLAs.

In response to these challenges, our previous work \cite{laclau_enhancing_2024} proposed to define a user experience (UX) focused runtime priority for non-safety-critical services. We introduced the concept of Automotive eXperience Integrity Level (AXIL), illustrated in
Table \ref{tab:axil}, to express user preferences and establish connections with SLAs. This metric allowed us to define a heuristic algorithm running onboard to orchestrate applications while optimizing the \textit{UX-to-resource} ratio of user-focused features. AXIL is similar to the already well-established ASIL levels \cite{frigerio_component-level_2019}, which help engineers optimize the \textit{safety-to-cost} ratio of safety-critical hardware and software components.
In our approach, applications declare runtime modes with varying feature sets, static resource requirements (e.g. CPU, memory, network bandwidth, energy consumption), and associated AXIL scores. When the requested features by the user exceed the available resources, the algorithm selects a mode for each application to maximize the overall UX. This approach allows the vehicle to adapt its current state to the current context (i.e. resources and user requests) while respecting the vehicle SLAs.

Our previous work \cite{laclau_enhancing_2024} demonstrated the effectiveness of this approach through simulation results. However, a comprehensive performance assessment would greatly contribute to validating its industrial feasibility. In this current work, we aim to contribute to the field by presenting (1) a dedicated test bench mimicking a typical SDV architecture, (2) realistic test scenarios and validation criteria such as resource usage and network health metrics, and (3) performance results.

The remainder of this paper is organized as follows. Section \ref{sec:background} presents the related work. Section \ref{sec:problem} summarizes the problem statement and solution presented in \cite{laclau_enhancing_2024}. Section \ref{sec:setup} presents the experimental setup and Section \ref{sec:results} discusses the experimental results. Section \ref{sec:conclusion} concludes the paper.

%% file: sections/02_background.tex
\section{Related work} \label{sec:background}

The traditional static allocation of resources for onboard applications in vehicles, accompanied by pre-defined rule-based local state management strategies, has long been the norm to ensure a stable and safety-certified software stack \cite{kugele_service-orientation_2017}. However, the advent of dynamic use cases and updates challenges the feasibility of maintaining control over the combinations of active applications requested by users, vehicles, or road contexts \cite{schindewolf_comparison_2022}. The conventional rule-based approach falls short in addressing the evolving network traffic requirements to apply context-aware degraded states.

Recognizing this limitation, the automotive industry is transitioning towards dynamic onboard service orchestration leveraging Service Oriented Architectures (SOA) capabilities. This paradigm shift aims to optimize resource usage and reduce engineering efforts for state management \cite{frtunikj_run-time_2014}. The incorporation of SOA facilitates vehicle-wide monitoring for centralized and context-aware global orchestration \cite{liao_cooperative_2022}.

Key to this evolution is the utilization of reconfigurable technologies such as Time-Sensitive Networking (TSN) and Software-Defined Networking (SDN) within the in-vehicle network. These technologies enable dynamic changes in the behavior and resource allocations of onboard software components through reconfigurations \cite{hackel_multilayered_2023,shi_recent_2023,halba_robust_2018}. Research initiatives propose atomic vehicle-wide reconfiguration strategies employing TSN and SDN to ensure safe transitions, even in dynamic driving scenarios \cite{hackel_software-defined_2019,kampmann_dynamic_2019,haeberle_softwarization_2020}. Examples of global mechanisms enabled by these strategies include improved network intrusion detection \cite{hackel_multilayered_2023}. As a result, the focus shifts from manually generating local state management rules to designing a centralized autonomous state management stack.

Additionally, recent research considers bringing containerization to automotive. Containers package applications and their dependencies into an OS-level isolated namespace to achieve hardware-software decoupling and maximum portability. Cloud platforms then distribute containers into clusters of servers according to workload specifications. Dynamic orchestration mechanisms then continuously reconfigure allocations to accommodate for varying workload demand~\cite{truyen_comprehensive_2019}.

While current commercial vehicles already implement static containers, applying this additional dynamic layer could bring enhanced use cases, e.g. switching between "autonomous" and "parked" allocations to maximize hardware utilization at all times. However, safety and embedded constraints must be addressed before industrial implementation.

Vehicles host two categories of applications: safety-critical (SC) services and best-effort (BE) applications. SC services require QoS guarantees, including real-time computing and time-sensitive network flows with predefined latency, jitter, and isolation requirements \cite{peng_survey_2023,kulzer_novel_2020}. BE applications, designed to enhance user experience, may function with varying resource allocations depending on the available resources. Each can define their own isolation requirements. 

While existing work has investigated dynamic container performance using automotive \cite{hackel_multilayered_2023} or general \cite{fernandez_blanco_can_2023} hardware, we have not found proposals for resource allocation strategies.
In this work, we focus on studying the experimental validity of onboard resource allocation built on top of the previously mentioned state-of-the-art dynamic mechanisms.

%% file: sections/03_problem.tex
\section{Methodology} \label{sec:problem}

This section serves as a concise summary of the problem formulation presented in our previous paper \cite{laclau_enhancing_2024}. We introduce the problem assumptions, goals, and modelling, followed by a brief overview of our proposed algorithm.

\subsubsection{Problem formulation}

Our approach starts with some assumptions and goals. We assume that each application is already assigned to a specific software context or ECU within the vehicle. The main goal of the algorithm is to continuously maximize the overall vehicle-wide user experience. We defined an algorithm that can make fast decisions in an embedded context, adjusting onboard functionalities when resource limitations hinder the allocation of the requested~resources.

Let $\mathcal{A}$ be the set of available applications in a vehicle app store. We separate $\mathcal{A}$ in two partitions, namely $\mathcal{A}_{SC}$ for safety-critical (SC) applications and $\mathcal{A}_{BE}$ for best-effort (BE) applications, with $n_{\text{SC}} = |\mathcal{A}_{\text{SC}}|$ and $n_{\text{BE}} = |\mathcal{A}_{\text{BE}}|$. We assume that SC applications are not subject to the same resource constraints as BE applications, as their resources are guaranteed to be reserved by the vehicle architecture separately. This may be performed by methods such as \cite{laclau_predictive_2023} based on pre-allocations which are out of the scope of this work. In this setup, the vehicle may dynamically change the onboard resource allocations depending on the required SC features. Hence, resources for BE applications must adapt at runtime depending on the current context, such as V2X network saturation and available BE network bandwidth.

Each BE application $\mathcal{A}_i \in A_{BE}$ defines a set of $m_i$ runtime modes, denoted as $\mathcal{A}_i^1$ to $\mathcal{A}_i^{m_i}$, with varying levels of functionalities and resource requirements. $\mathcal{A}_i^1$ is the nominal mode with the most features and resource requirements, while $\mathcal{A}_i^{m_i}$ is the most degraded mode. Modes can have dependencies to other modes, forming a directed acyclic~graph.

The vehicle can be modelled as a list of $r$ resources $\mathcal{R}_1, \ldots, \mathcal{R}_r$. The maximum hardware capacity of the vehicle is denoted as a vector of $r$ positive reals $R^{\text{max}}$ such that $R^{\text{max}}[i] \in \mathbb{R}^+$ denotes the maximal capacity of resource $\mathcal{R}_i$. Resources can describe the CPU and memory capacities for each ECU, available best-effort bandwidth for each physical network link in each direction, external system capacity such as V2X and edge computing availability, and more.
However, resource availability can dynamically change at runtime. Hence, we define the current BE capacity of a vehicle as a vector $R$ of $r$ values such that $R[i] \in [0, R^{\text{max}}[i]]$.

\input{sections/90_table.tex}

\begin{figure}[t!]
    \centering
    \includegraphics[clip, trim=0cm 0cm 0cm 0cm, width=\linewidth]{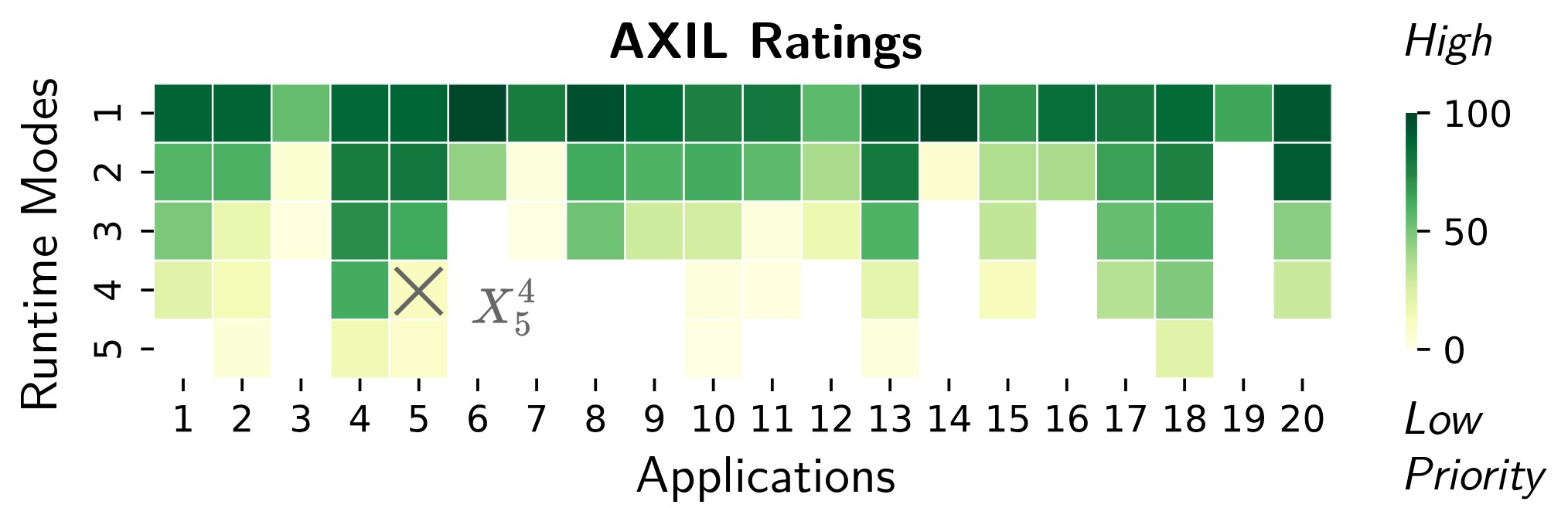}
    \caption{Example of AXIL ratings attributed to the 1-5 runtime modes for each of the 20 applications. Higher modes provide more features, hence better QoE and execution priority.}
    \label{fig:instance}
\end{figure}

Each mode $\mathcal{A}_i^j$ is associated with an AXIL rating $X_{i}^j \in \mathbb{R}^+$, expressing the perceived user experience for the application in this mode. The higher the AXIL rating, the better the user experience. Figure \ref{fig:instance} illustrates an example of randomly selected AXIL ratings for a set of applications. This creates a fictional but representative desired state instance. Additionally, each mode $\mathcal{A}_i^j$ is associated with a resource requirements vector $M_{i}^j$, with $M_{i}^j [k] \in [0, R^{\text{max}}[k]]$ expressing the resource requirement of resource $\mathcal{R}_k$ to execute the mode. At any time, the vehicle may request any subset of~$\mathcal{A}$.

At the core of the optimization problem is selecting the best combination of runtime modes for a given set of requested applications, with some applications potentially remaining disabled. The objective is to maximize overall user experience, defined as the sum of the AXIL scores of each active mode, while respecting resource limits and dependency relationships. As shown in our previous work, this translates into a mathematical problem comparable to the NP-Hard \textit{knapsack} problem, requiring a heuristic solution.

\subsubsection{Proposed algorithm}

We proposed a heuristic solution for efficient runtime mode selection. In essence, the algorithm starts with all applications disabled and iteratively enables the most beneficial modes, starting from the most degraded mode upward, until the resource limits are reached. At each iteration, the next higher mode for each application is considered as a potential candidate to be upgraded. Then, one of the candidates is selected using a cost function, which divides the AXIL improvement by the resource cost of enabling each mode. The algorithm upgrades the selected mode and continues iterating until there is no improvement left. The algorithm also respects the dependencies between the modes, ensuring that a mode is only enabled if all its dependencies are also enabled. This approach is designed to be embedded onboard, making suboptimal but fast decisions in real-time. It can be re-executed when dynamic changes in resource availability and user requests occur to continuously adapt the onboard features.

While our previous simulation results demonstrated its effectiveness, an experimental study would greatly contribute in validating its industrial feasibility, notably using embedded hardware with capabilities resembling automotive ECUs to study transition delays and system reactivity.

%% file: sections/90_table.tex
\begin{table}[t!]
    \caption{AXIL definition. Just like ASIL, AXIL combines three parameters to assess the runtime priority of a service.}
    \centering

    \begin{tabular}{l|l||cccc}
        \multicolumn{1}{x{0.9cm}|}{\mr{2}{$E_1$}} & \multicolumn{1}{x{0.9cm}||}{\multirow{2}{*}{$E_2$}} & \multicolumn{4}{c}{$E_3$}                                                                                                         \\
                                                  &                                                     & \multicolumn{1}{x{1cm}}{Minimal} & \multicolumn{1}{x{1cm}}{Low} & \multicolumn{1}{x{1cm}}{Medium} & \multicolumn{1}{x{1cm}}{High} \\
        \hline
        \mr{4}{Easy}                              & Rare                                                & -                                & -                            & -                               & -                             \\
                                                  & Low                                                 & -                                & -                            & -                               & -                             \\
                                                  & Medium                                              & -                                & -                            & -                               & A                             \\
                                                  & High                                                & -                                & -                            & A                               & B                             \\
        \hline
        \mr{4}{Medium}                            & Rare                                                & -                                & -                            & -                               & -                             \\
                                                  & Low                                                 & -                                & -                            & -                               & A                             \\
                                                  & Medium                                              & -                                & -                            & A                               & B                             \\
                                                  & High                                                & -                                & A                            & B                               & C                             \\
        \hline
        \mr{4}{Difficult}                         & Rare                                                & -                                & -                            & -                               & A                             \\
                                                  & Low                                                 & -                                & -                            & A                               & B                             \\
                                                  & Medium                                              & -                                & A                            & B                               & C                             \\
                                                  & High                                                & A                                & B                            & C                               & D                             \\
    \end{tabular}

    \vspace{1mm}
    \hspace{25mm} \textit{Priority order: - $<$ A $<$ B $<$ C $<$ D }

    \vspace{6mm}

    \begin{tabular}{c|c|c|c} 
        Legend $\rightarrow$ & $E_1$                & $E_2$              & $E_3$           \\
        \hline
        \textbf{ASIL}        & Controllability      & \mr{2}{Exposition} & Severity        \\
        \textbf{AXIL}        & Ease of Substitution &                    & Quality of Exp. \\
    \end{tabular}

    \label{tab:axil}
\end{table}

%% file: sections/04_setup.tex
\section{Experimental setup} \label{sec:setup}

This section presents the experimental setup used to evaluate our proposal. We describe our design choices for the test bench, the hardware used, the software architecture, the test scenarios, the data collection, and the evaluation metrics.

\subsubsection{Objectives}

In this study, we aim to provide empirical evidence of the benefits brought by our approach through experimental results. Hence, we focused on building a general, flexible, and representative platform rather than using specialized automotive hardware. This reduces development time and offers greater flexibility for implementations and scenarios, while still providing a realistic environment for the algorithm to be evaluated, as justified in the following paragraphs. Additionally, as mentioned in the related works, existing automotive platforms have already been used to investigate container performance \cite{hackel_multilayered_2023} which is not our~scope.

\subsubsection{Hardware architecture}

\begin{figure}[t!]
    \centering
    \includegraphics[clip, trim=0cm 0cm 0cm 0cm, width=\linewidth]{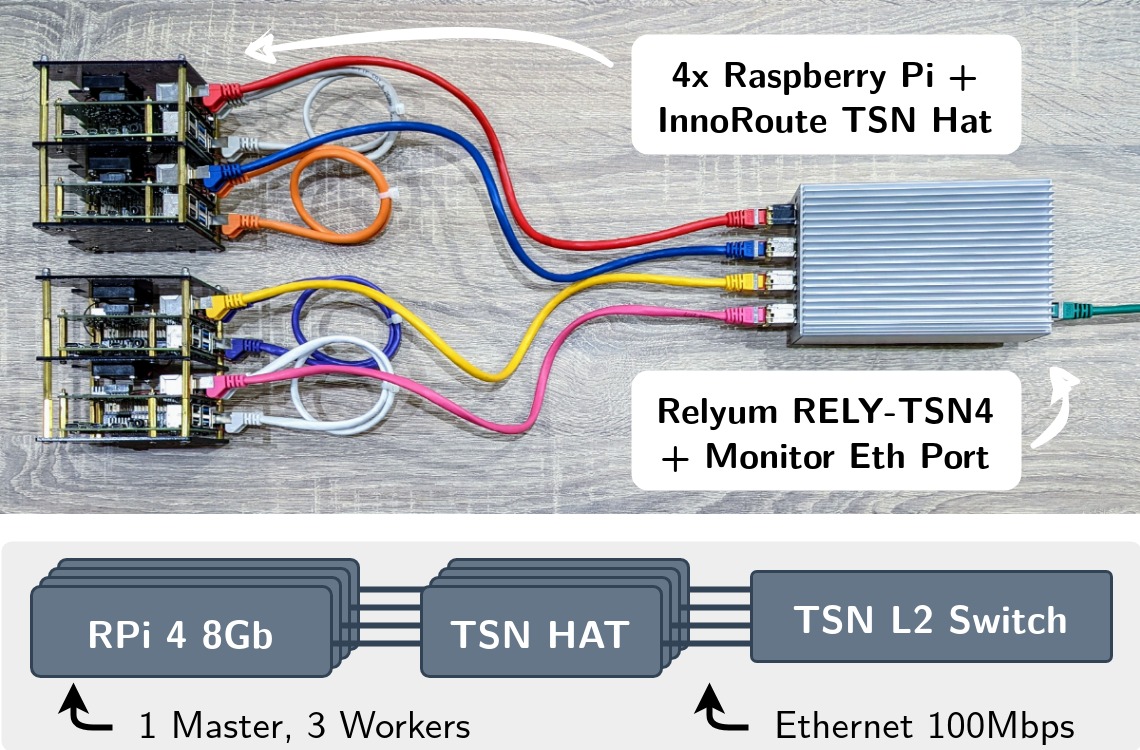}
    \caption{Hardware architecture of test bench made of 4 ECUs connected in star topology using TSN-enabled network links.}
    \label{fig:bench}
\end{figure}

The test bench is designed to resemble a typical SDV architecture. It is composed of a set of 4 ECUs connected in a star topology using Ethernet links. The ECUs are represented using Raspberry Pi 4 Model B (RPi) single-board computers with 8Gb of RAM, connected through an Ethernet switch. As SDVs are expected to include mixed-criticality network traffic, all network interfaces are TSN-capable using an extension card for the RPis and TSN features in the switch model Relyum RELY-TSN4. See Figure \ref{fig:bench} for a photo and simplified representation of the test bench. The RPis are connected to an external computer for monitoring through Wi-Fi, which enables network flow separation between scenario-relevant and monitoring traffic.

\subsubsection{Software architecture} \label{sec:setup-software}

The software stack is designed to resemble a typical SDV software architecture. Each RPi runs a Linux-based operating system patched with a real-time kernel to enable TSN capabilities. The network stack supports the TSN standards including time synchronization (IEEE 802.1AS) and time-aware traffic shaping (IEEE 802.1Qbv). Finally, all ECUs and switch have an active local server supporting the NETCONF protocol, allowing for dynamic reconfiguration of the TSN ports. Hence, a centralized controller can dynamically change the TSN configuration of the network to adapt to the current vehicle context by calling the NETCONF server on each device with a precise time.

\begin{figure*}[t!]
    \centering
    \includegraphics[clip, trim=0cm 0cm 0cm 0cm, width=\linewidth]{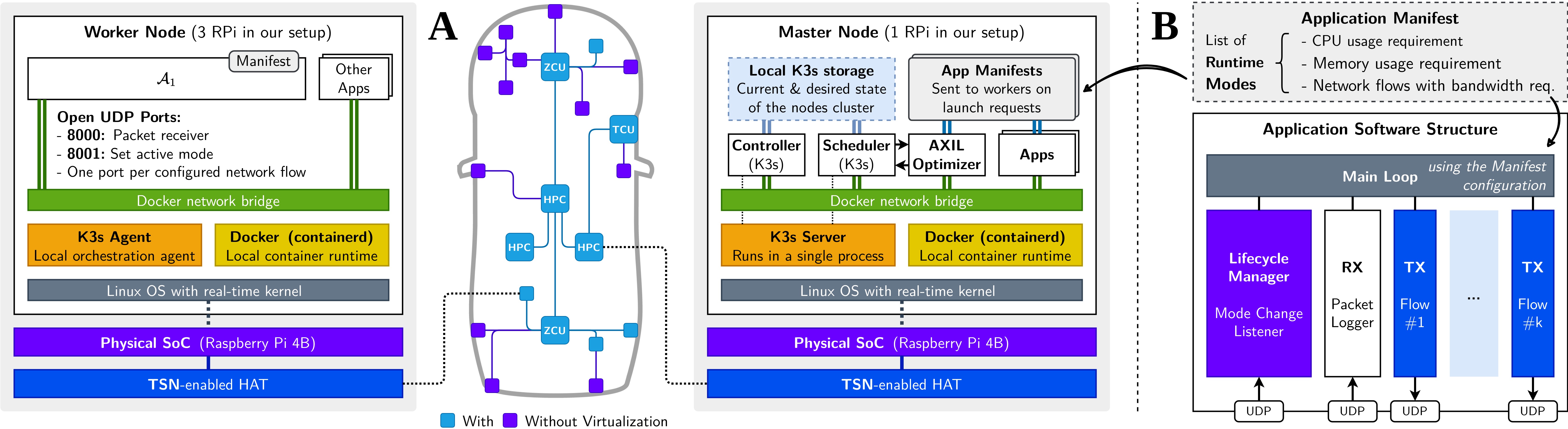}
    \caption{Simplified representation of the software stack. \textbf{(A)} We use K3s as the Kubernetes distribution to orchestrate and distribute the containerized applications and their runtime modes into the ECUs. \textbf{(B)} The application is structured to generate mock best-effort network traffic according to a manifest file. These manifests are distributed to the ECUs at launch time.}
    \label{fig:arch}
\end{figure*}

Figure \ref{fig:arch} shows a component view of the software stack. We use K3s as the Kubernetes distribution to orchestrate and distribute the applications and their runtime modes into the ECUs. This corresponds to the future trend of automotive software development for SDVs \cite{nayak_automotive_2023}, which aims to facilitate software-hardware decoupling, continuous deployment, and dynamic resource allocation. Hence, the applications are containerized using Docker \cite{fernandez_blanco_can_2023}. They are all launched from a common image, but they are configured with different runtime modes, network traffic generation profiles, and resource requirements from a manifest file. The applications are designed to generate best-effort network traffic and artificially consume resources following pre-determined requirements, such as CPU and memory usage. With this approach, we aim to mimic the typical resource constraints and behavior found in BE applications.

\subsubsection{Configuration generation}

Given a number of applications $n$ and maximum number of runtime modes $m_{\text{max}}$, we randomly generate a number of modes $m_i$ for each application $\mathcal{A}_i$.
We generate a random dependency graph $G_{\mathcal{A}}$ between applications with a target density by iteratively adding edges randomly to a graph up until the density is reached, and removing one edge per cycle if any appears. This lets us generate a second mode-level dependency graph $G_{\text{M}}$. For each dependency edge in $G_{\mathcal{A}}$, we generate a random number of edges between modes of the two apps as long as they do not cross. Then, for each edge in $G_{\text{M}}$, we generate a random number of network flows with random bandwidth requirements. Finally, each mode $\mathcal{A}_i^j$ is attributed a random resource requirement $M_i^j [k]$ for each resource $k$, i.e. CPU and RAM in this work. Note that values are generated within the bounds set in Table \ref{tab:ranges} for fixed values. They also respect the structure defined in Section \ref{sec:problem} along with its constraints.

We generate a manifest file for each application $\mathcal{A}_i$ which describes the runtime modes, network flows, and resource requirements for each mode $\mathcal{A}_i^j$. When an application is launched in a specific mode, the common container image is started and configured with the corresponding manifest. The selected value ranges for each requirement have been manually calibrated relative to the test bench capabilities, as the aim is to demonstrate the ability of the algorithm to adapt to the current platform. These ranges are shown in Table~\ref{tab:ranges}.

Then, each mode is assigned a random AXIL rating in decreasing order, as illustrated in Figure \ref{fig:instance} with a particular instance of 20 applications specifying at most 5 runtime modes each. Finally, each application is randomly assigned to one of the ECUs in the manifest, which we assume is provided by an external onboard scheduler out of our scope.

\subsubsection{Test scenario}

We define a simple yet representative test scenario to evaluate our algorithm. We aim to let the test bench run through a continuous change of vehicle states corresponding to subsets of active applications in the app store. To generate a state, a random set of applications is added and recursively extended with their dependencies. Then, we evaluate the test bench performance through continuous state changes, each with a random duration of 10 to 60 seconds.

As we have found the computing power of the RPis to be limited, the network bandwidth requirements are set to be relatively low. We set a 90\% TAS closed duty cycle on all TSN ports, which effectively limits the physical bandwidth to 10Mbps. This setting can also be seen as if safety-critical applications have already reserved network resources for time-sensitive traffic. Therefore, the available bandwidth given to our optimization algorithm is 10Mbps per link to reflect the current available BE network state in the vehicle.

We repeat this scenario in two settings, (1) by activating all applications at their maximum runtime mode thereby bypassing the algorithm (baseline), and (2) by launching the applications at the runtime mode selected by the algorithm (optimized). This allows us to compare the utility of the algorithm compared to the lack of a resource-aware mechanism.

\begin{table}[t!]
    \centering
    \normalsize
    \caption{Problem parameter ranges for the test scenarios.}
    \label{tab:ranges}
    \begin{tabular}{p{6cm}|c}
        \textbf{Resource}                      & \textbf{Value range} \\ \hline
        CPU usage                              & 0-10\%               \\
        Memory usage                           & 0-200Mb              \\
        Number of modes                        & 1-5                  \\
        Number of flows per dependency         & 1-5                  \\
        Network bandwidth requirement per flow & 0.1-2Mbps            \\
    \end{tabular}
\end{table}

\subsubsection{Evaluation metrics}

We consider several evaluation metrics to assess the performance of the algorithm:

\begin{itemize}
    \item \textbf{Network health}: As the main objective of our algorithm is to guarantee the allocation of resources specified by each application's manifest, our primary metric aims to measure the overall system's network health. We measure the network traffic generated and received by the applications. For each flow, we compare the observed network bandwidth with the expected generated bandwidth specified in its manifest file. This results in a time series of percentage values. We then aggregate the results with the median, Q1, and Q3 values as the indication of global vehicle health.

    \item \textbf{Resource usage}: We also measure the resource usage of the ECUs, namely CPU and memory usage, to assess the ability of the algorithm to respect the resource constraints of the vehicle. Observing a low network health along with saturated computing nodes would signal the necessity of resource-aware orchestration mechanisms.

    \item \textbf{Algorithm performance}: At each vehicle state change, we measure the time taken by our algorithm to generate a solution. Additionally, we also monitor the time to reconfigure the K3s cluster once a new state is requested, and the time to adapt to dynamic changes in resource availability and user requests. This will provide insights into the performance of this architecture paradigm. Note that the total user experience is not directly measured in this study. We consider the algorithm always returns its best approximate solution as demonstrated in our previous work \cite{laclau_enhancing_2024}.
\end{itemize}

\subsubsection{Data collection}

While a scenario is running, the applications store the packets received, resource usages, and notable lifecycle events such as activation times and mode changes. They are then post-processed to extract the metrics.

%% file: sections/05_results.tex
\begin{figure}[b!]
    \centering
    \begin{tabular}{l||c|c|c}
        \textbf{Name} & \textbf{Nb. of Apps} & \textbf{Nb. of Modes} & \textbf{Dependencies} \\ \hline
        XS            & 10                   & 1                     & 5\%                   \\
        S             & 20                   & 3                     & 5\%                   \\
        M             & 30                   & 4                     & 10\%                  \\
        L             & 50                   & 5                     & 15\%                  \\
        XL            & 100                  & 5                     & 20\%                  \\
    \end{tabular}

    \vspace{3mm}

    \includegraphics[clip, trim=0.3cm 0.2cm 0cm 0cm, width=\linewidth]{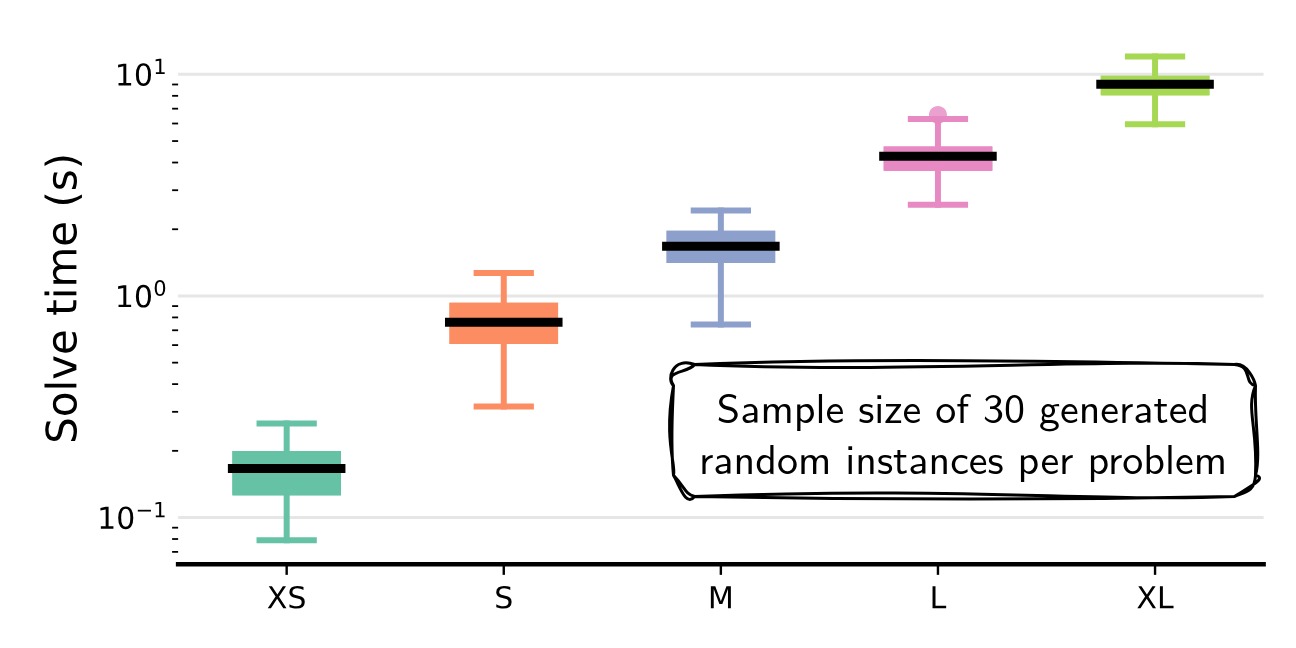}
    \caption{Solving times on the test bench depending on the problem size with parameters given in the associated table.}
    \label{fig:times}
\end{figure}

\section{Results} \label{sec:results}

\begin{figure*}[t!]
    \centering
    \includegraphics[clip, trim=0.2cm 0cm 0.4cm 0cm, width=\linewidth]{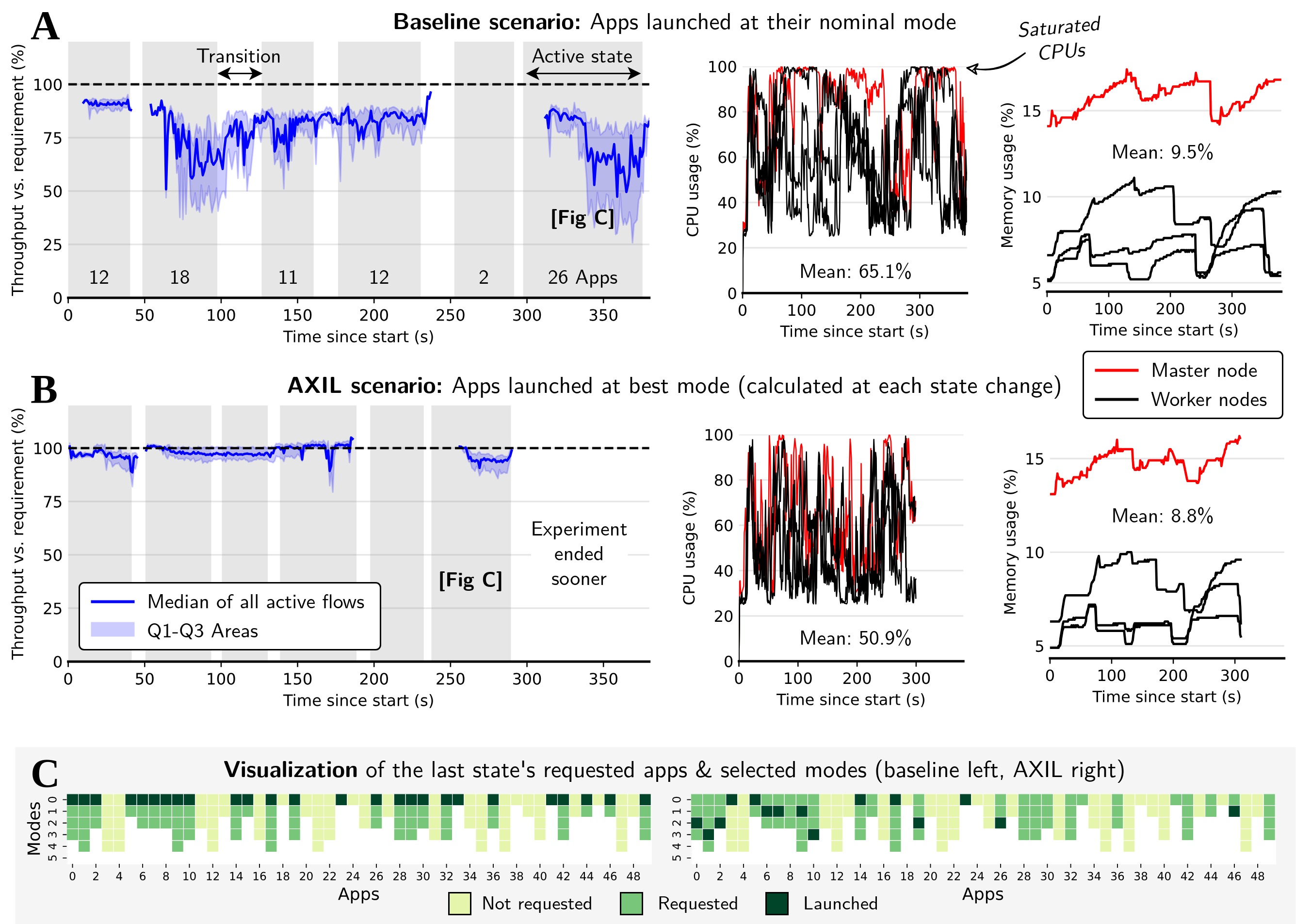}
    \caption{Experimental results with an app store of 50 applications and 6 state changes of random sets of applications. We show our proposed network health metric when launching apps at \textbf{(A)} their nominal modes and \textbf{(B)} at the mode calculated by our algorithm. \textbf{(C)} Illustrates the runtime mode decisions in both scenarios with the requested applications and launched~modes.}
    \label{fig:results}
\end{figure*}

The experimental results are presented in Figures \ref{fig:times} and~\ref{fig:results}. The results demonstrate that the baseline scenario shows severe stress in CPU usage, and the applications are not able to send network traffic at their target speeds. On the opposite, the optimized scenario chooses to launch the requested applications at degraded states to accommodate for the currently available resources, finding a suitable UX compromise.
In this section, we start by studying the algorithm performance then analyze a full test scenario with random state changes.

\subsubsection{Adaptability in real-time}

On this hardware, the algorithm search times are shown in Figure \ref{fig:times}. It is capable of producing solutions in 750ms to 2.6 seconds for approximately 30 applications with 1 to 4 modes each (size M). We believe this problem size to be a common use case in~commercial vehicles, e.g. to manage user-focused or optional services.

This performance is suitable for occasional best-effort decision-making during vehicle operation, and can be re-executed when dynamic changes in resource availability and user requests occur. However, limitations remain as the algorithmic complexity prohibits problem sizes larger than M in practice. Further optimizations can be achieved such as reimplementing the algorithm in a more efficient language (currently in Python), caching common results, pre-activating some modes depending on global state management rules, or performing calculations on the edge or cloud when available.

\subsubsection{Network performance}

Figure \ref{fig:results} shows the metrics collected from an experiment with an app store of size M. It compares the worst-case scenario where all applications are activated at their maximum runtime mode (Figure \ref{fig:results}A) with the scenario where our algorithm is active (Figure \ref{fig:results}B). The results show that the algorithm effectively respected the constraints of the vehicle as the network health score remains close to 100\%. Hence, applications are degraded to their most efficient mode. The baseline shows an unpredictable and insufficient network performance due to over-provisioned resources, demonstrating the need of using resource-aware mechanisms. Instead, we maximize the UX-to-resource ratio.

Therefore, only the optimized scenario with 100\% network health permits the applications to obtain their required network resources specified in their manifests. Note that container launch delays explain temporary missing network statistics, and the 5th state includes 2 applications installed on the same node. Their traffic is not seen in our Ethernet-scoped statistics but is accounted for in the compute usage.

\subsubsection{Optimized resource usage}

The baseline scenario shows that the CPU usage of nodes regularly reaches saturation at 100\%, also indicating over-provisioning. In the optimized case, only the master node rarely reaches CPU saturation mostly during transitions (i.e. scheduling, AXIL calculation, deployment). This behavior is desired in automotive systems to avoid performance bottlenecks and can be calibrated through the resource usage parameters. Memory usage is comparable in both scenarios, which could be due to the small meory footprints of the applications.

\subsubsection{Reduced transition times}

Figure \ref{fig:results} also shows a significant decrease in total transition times between vehicle states when the algorithm was applied. The transition times include stopping and launching the application containers, as well as the algorithm search time in the optimized case. This is a direct result of using the available resources without overloading the system, which enables faster operations for K3s. The algorithm does not significantly impact the total transition times in this setup, as the search time is negligible compared to the container launch times.

The overall results lay a first stone for advancing dynamic onboard service orchestration in SDVs.
However, we must note several limitations to our results. First, applications are implemented in Python, which introduces significant performance bottlenecks and may not reflect the maximal performance reachable with this hardware. This prohibited the study of packet latencies and jitter, as well as including time-sensitive network support for applications. Second, the current work does not dynamically change the available resources such as network capacity which limits the relevance 

\noindent of the results to real-world scenarios.

Limitations also apply to the proposed resource-aware dynamic methodology. First, the complexity of the algorithm prohibits large sets of requested applications. The computation time remains too high for time-critical decisions, and the algorithm has no real-time guarantees. This may limit use cases as many onboard operations must be performed within strict global timing requirements. Second, this dynamic management approach is only valid for microprocessor-based environments, as it requires K3s and Docker. Finally, the algorithm is limited by the current BE resources available on the vehicle and has no control over allocations for other domains such as SC apps. However, it presents an opportunity to investigate the deployment of more sophisticated SC applications within complex ECUs, such as High Performance Computer (HPC) ECUs, where an increasing number of safety, body, and infotainment applications are being allocated using shared resources.

%% file: sections/10_conclusion.tex
\section{Conclusion} \label{sec:conclusion}

This paper introduces an experimental investigation into a heuristic algorithm designed for efficient resource-based dynamic application orchestration in SDVs. The algorithm prioritizes user experience by optimizing the selection of runtime modes of requested applications within resource constraints and dependency relationships set by developers.

Conducted on a dedicated test bench mimicking a typical SDV architecture, the study employs Raspberry Pi single-board computers connected via Ethernet supporting TSN standards. Results illustrate the algorithm's capability to continuously adapt onboard functionalities while keeping the resource usage within ECU and network capacities. The concept of considering runtime modes combined with this algorithm guarantees sane network health and resource usage, independently of the ever-changing user requests. It also decreases transition times between vehicle states, compared to launching all apps without using any resource control mechanism. In this setup, the optimization algorithm can perform decisions in approximately one second for 20 applications with 1-5 modes each, which is a promising result for continuous orchestration. Room for performance improvement remains if the presented system is reimplemented using embedded automotive-grade technologies.

These results lay a foundation for advancing dynamic onboard service orchestration in SDVs. Future work will focus on extending the study to more intricate scenarios and diverse applications, as well as including other application domains such as safety-critical and cooperative V2X services.

\section*{Acknowledgments}

This work was supported by Stellantis under the collaborative CIFRE framework UTC/CNRS/PCA (ANRT contract n°2021/0865) with Heudiasyc. The authors would like to thank Charles Perold for his technical contributions.